\documentclass[prd,aps,twocolumn,nofootinbib]{revtex4}

\usepackage{graphicx}

\begin{document}

\title{Gravitino dark matter and baryon asymmetry from $Q$-ball decay in gauge mediation}

\author{Shinta Kasuya$^a$ and Masahiro Kawasaki$^{b,c}$}

\affiliation{
$^a$ Department of Information Sciences,
     Kanagawa University, Kanagawa 259-1293, Japan\\
$^b$ Institute for Cosmic Ray Research,
     University of Tokyo, Chiba 277-8582, Japan\\
$^c$ Institute for the Physics and Mathematics of the Universe, 
     University of Tokyo, Chiba 277-8582, Japan}

\date{July 2, 2011}

\begin{abstract}
We investigate the $Q$-ball decay in the gauge-mediated SUSY breaking.
$Q$ balls decay mainly into nucleons, and partially into gravitinos, while they
are kinematically forbidden to decay into sparticles which would be cosmologically 
harmful. This is achieved by the $Q$-ball charge small enough to be unstable for 
the decay, and large enough to be protected kinematically from unwanted decay channel.
We can then have right amounts of the baryon asymmetry and the dark matter of the
universe, evading any  astrophysical and cosmological observational constraints
such as the big bang nucleosynthesis, which has not been treated properly in the
literatures.
\end{abstract}



\maketitle

\section{Introduction}
Production of dark matter and baryon number of the universe is very important 
in cosmology. In the gauge-mediated supersymmetry (SUSY) breaking, the gravitino 
is natural candidate of the dark matter, while the Afflick-Dine (AD) baryogenesis \cite{AD} 
is promising mechanism because thermal leptogenesis confronts serious 
cosmological gravitino problem, for example. It is well known that the AD condensate
transforms into nontopological solitons, $Q$ balls, during the course of baryogenesis 
\cite{KuSh,EnMc,KK1,KK2,KK3}.
The charge of the $Q$ ball in the gauge mediation could be large enough to be stable
against the decay into nucleons, and the $Q$ ball itself could be the dark matter of 
the universe \cite{KuSh}. In this case, however, nondetections in direct and/or indirect 
$Q$-ball searches tell us that it may be difficult to simultaneously obtain enough 
baryon asymmetry of the universe\footnote{%
More recent observational limits tighter than those applied in Ref.\cite{KK3} may
wipe out the allowed regions.}. 

On the other hand, the small charge $Q$ ball will decay into nucleons. 
At the same time, it could decay into gravitinos, which would have enough abundance
to be the dark matter. This idea was considered in Refs.\cite{ShKu,DoMc}\footnote{
Its possibility was mentioned in Ref.\cite{KK3}.}.
In the former, the authors considered relatively small charge $Q$ balls which evaporate
in the thermal bath, and imagined that the gravitino is directly produced from the $Q$ balls. 
In this case, however, most of the charge decays (evaporates) at the temperature around 
the electroweak scale, and sparticles are created, finally leading to the production of the 
next-to-lightest SUSY particle (NLSP), which thermalizes immediately. The gravitino 
could be produced by the NLSP decay, but the abundance of the gravitino is too small 
to be the dark matter of the universe, and the NLSP decay would destroy light elements 
during the big bang nucleosynthesis (BBN).

In the latter reference, the formed $Q$ ball is the gravity-mediation type, which is able to
decay into NLSPs nonthermally. The NLSP decays into the gravitiono in the end, but they
did not discuss the fact that the NLSP abundance is limited severely by the BBN constraints
for $m_{3/2} \gtrsim 1$ GeV,  so that the amount of the produced gravitino should be much 
smaller than that to be the dark matter \cite{BBN} \footnote{
Their model may work for a right-handed sneutrino LSP or if there is a small amount 
of R-parity violation\cite{McDonald}.
}.

In this article, we investigate a very simple scenario that the unstable $Q$ balls decay 
mainly into nucleons, partially into gravitinos with small branching ratio, while
kinematically forbidden to NLSPs almost throughout the decay process, in the gauge-mediated
SUSY breaking. The crucial observation reveals that the right amounts of the baryon asymmetry
and the dark matter of the universe can be created through this process. 
The scenario is achieved by a couple of key ingredients. One is that the $Q$ ball 
decays from its surface \cite{evap} so that the decay rate through the main channel has an
upper bound, which is called the saturated rate, while the decay into gravitino is not 
saturated. Another is that a large enough $Q$ ball forbids the decay into the NLSP, while 
allowing the decay into nucleons and gravitinos, having an appropriate lifetime to decay 
at the temperature of $O$(10 MeV).

The whole scenario proceeds as follows. The AD field starts the oscillation (rotation) 
after inflation. The rotation has an oblate orbit in the potential. Due to the negative pressure,
the AD field feels spatial instabilities, which grow into $Q$ and anti-$Q$ balls. The $Q$
and anti-$Q$ balls decay into nucleons and anti-nucleons to create the baryon asymmetry
in the universe, while producing small amount of gravitinos which become the dark matter
of the universe. NLSPs would be produced only at the very end of the decay process 
when the charge of the Q ball becomes small enough to open the kinematically allowed 
channel to the NLSP production\cite{KT}, and its abundance is small enough to evade any 
cosmological disasters such as spoiling the successful BBN.

The structure of the article is as follows. In the next section, we show the features of the
$Q$ ball in the gauge-mediated SUSY breaking model, and the details of the decay
process of the $Q$ ball are described in Sec.~\ref{Qdecay}. In Sec.\ref{abundances},
we estimate the abundances of the baryon number, the dark matter, and the NLSP. 
We impose BBN constraints on the NLSP abundances to derive the allowed gravitino
mass range in Sec.\ref{BBNconstraints}. In Sec.\ref{scenario}, we show the region in the
model parameter space where the successful scenario exists. Section \ref{conclusion} is
devoted to our conclusions.

\section{Q ball in gauge mediation}
The AD field consists of some combination of squarks and sleptons. The potential is flat
in the SUSY-exact limit, but due to the breaking of SUSY in gauge mediation, it is 
lifted as\cite{log2}
\begin{equation}
V(\Phi)=\left\{
\begin{array}{ll}
m_\phi^2|\Phi|^2, & (|\Phi| \ll M_S) \\
M_F^4 \left( \log \frac{|\Phi|^2}{M_S^2} \right)^2, & (|\Phi| \gg M_S)
\end{array}\right.,
\end{equation}
where $m_\phi \sim O({\rm TeV})$ is a soft breaking mass, and $M_F$ and $M_S$ 
are related respectively to the $F$ and $A$ components of a gauge-singlet chiral 
multiplet $S$ in the messenger sector as
\begin{equation}
M_F^4 \equiv \frac{g^2}{(4\pi)^4}\langle F_S \rangle^2, \qquad 
M_S\equiv \langle S \rangle.
\end{equation}
Here $g$ generically stands for the standard model gauge coupling, and $M_F$ is
allowed in the following range:
\begin{equation}
10^3 \ {\rm GeV} \lesssim M_F \lesssim \frac{g^{1/2}}{4\pi}\sqrt{m_{3/2}M_{\rm P}},
\label{MFrange}
\end{equation}
where $m_{3/2}$ is the gravitino mass and $M_{\rm P} = 2.4\times 10^{18}$ GeV is the
reduced Planck mass.

The AD field starts the oscillation when $H \sim M_F^2/\phi_{\rm osc}$, where
$\phi_{\rm osc}$ is the field amplitude at the onset of the oscillations. The field fluctuations
grow exponentially due to the negative pressure during the helical motion, and the field 
transforms into $Q$ balls. The charge of the formed $Q$ ball is estimated as\cite{KK3}
\begin{equation}
Q=\beta \left(\frac{\phi_{\rm osc}}{M_F}\right)^4,
\label{Qosc}
\end{equation}
where $\beta\simeq 6\times 10^{-4}$ for a circular orbit ($\varepsilon=1$), while 
$\beta \simeq 6 \times 10^{-5}$ for an oblate case ($\varepsilon \lesssim 0.1$). 
Here $\varepsilon$ represents the ellipticity of the field orbit.
The charge $Q$ is just the $\Phi$-numbers, and relates to the baryon number
of the $Q$ ball as
\begin{equation}
B=b \, Q,
\end{equation}
where $b$ is the value of the baryon number carried by a $\Phi$ particle. 
For example, $b=1/3$ for the $udd$ direction.
The mass, the size, the rotation speed of the field, and the field value at the center
of the $Q$ ball are related to the charge $Q$ as 
\begin{eqnarray}
M_Q & \simeq & \frac{4\sqrt{2}\pi}{3} M_F Q^{3/4}, \\
\label{RQ}
R_Q & \simeq & \frac{1}{\sqrt{2}} M_F^{-1} Q^{1/4}, \\
\label{omegaQ}
\omega_Q & \simeq & \sqrt{2}\pi M_F Q^{-1/4}, \\
\label{phiQ}
\phi_Q & \simeq & M_F Q^{1/4},
\end{eqnarray}
respectively.

\section{Q-ball decay}
\label{Qdecay}
A $Q$-ball decay takes place if some decay particles carry the same kind of the 
charge of the $Q$ ball, and the mass of all the decay particle is less than $M_Q/Q$. 
In this case, the $Q$-ball decay rate $\Gamma_Q$ has an upper bound, which 
we call the saturated rate $\Gamma_Q^{\rm (sat)}$ \cite{evap}:
\begin{equation}
\Gamma_Q \lesssim \Gamma_Q^{\rm (sat)} 
\simeq \frac{1}{Q} \frac{\omega_Q^3}{192\pi^2} 4\pi R_Q^2
\simeq \frac{\pi^2}{24\sqrt{2}}M_F Q^{-5/4},
\end{equation}
where we use Eqs.(\ref{RQ}) and (\ref{omegaQ}) in the last equality. The decay rate
saturates typically when $f_{\rm eff} \phi_Q \gtrsim  \omega_Q$, where the elementary
process has an interaction of ${\cal L}_{\rm int}=f_{\rm eff} \phi \psi\bar{\psi}$. 
On the other hand, for $f_{\rm eff} \phi_Q \ll  \omega_Q$, the decay rate is given by\cite{evap}
\begin{equation}
\Gamma_Q \simeq 3\pi \frac{f_{\rm eff} \phi_Q}{\omega_Q} \frac{1}{Q}
\frac{\omega_Q^3}{192\pi^2} 4\pi R_Q^2
\simeq 3\pi \frac{f_{\rm eff} \phi_Q}{\omega_Q} \Gamma_Q^{\rm (sat)}.
\end{equation}

Since we are interested in the case that the $Q$ ball can decay into nucleons 
($M_Q/B>m_N$, where $m_N \simeq 1$ GeV is the nucleon mass) and, 
at the same time, the decay into NLSPs is kinematically prohibited 
($M_Q/Q  < m_{\rm NLSP}$), the main decay channel is the decay into nucleons.
We thus consider the $Q$-ball charge in the range $Q_{\rm cr} < Q < Q_{\rm D}$, where
\begin{eqnarray}
\label{Qcr}
& & Q_{\rm cr} = \frac{1024\pi^4}{81} \left(\frac{M_F}{m_{\rm NLSP}}\right)^4, \\
& & Q_{\rm D} = \frac{1024\pi^4}{81} \left(\frac{M_F}{b\, m_N}\right)^4.
\label{QD}
\end{eqnarray}
Notice that the NLSP is produced at the latest moment of the decay, only after the 
charge of the $Q$ ball becomes less than $Q_{\rm cr}$ during the course of 
the decay\cite{KT}.

The elementary process of the main channel is 
squark $+$ squark $\rightarrow$ quark $+$ quark via (heavy) gluino exchange, 
whose rate is estimated as\cite{Ellis}
\begin{equation}
\Gamma_\phi\simeq  \langle \sigma v \rangle n_\phi \simeq \frac{\zeta\alpha_s^2}{m_{\tilde{g}} 
\omega_Q} \omega_Q\phi_Q^2
\simeq \frac{1}{8\pi} \frac{\zeta g_s^4}{2\pi} \frac{\phi_Q^2}{m_{\tilde{g}}\omega_Q} \omega_Q,
\end{equation}
where $m_{\tilde{g}}$ is the gluino mass, and $\zeta \sim |V_{\rm CKM}|^4$ is a possible
CKM suppression factor ($10^{-3} \lesssim |V_{\rm CKM}| \lesssim 1$ \cite{PDG}). 
Thus, we obtain the effective coupling as
\begin{equation}
f_{\rm eff} \simeq \frac{\zeta^{1/2}g_s^2}{\sqrt{2\pi}} \frac{\phi_Q}{(m_{\tilde{g}}\omega_Q)^{1/2}}.
\end{equation}
Since we have
\begin{eqnarray}
& & \hspace{-5mm}
\frac{f_{\rm eff}\phi_Q}{\omega_Q} \simeq 1.8 \times 10^{20} \zeta^{1/2} g_s^2 
\left(\frac{M_F}{10^6~{\rm GeV}}\right)^{1/2} \nonumber \\
& & \hspace{20mm} \times 
\left(\frac{m_{\tilde{g}}}{\rm TeV} \right)^{-1/2}
\left(\frac{Q}{10^{23}}\right)^{7/8} \gg 1,
\end{eqnarray}
the main decay channel is saturated so that we must use the saturated rate 
$\Gamma_Q^{\rm (sat)}$ for the decay. Thus, the $Q$ ball decays when the cosmic
time becomes the lifetime of the $Q$ ball: $t \simeq 1/\Gamma_Q^{\rm (sat)}$. 
The $Q$ ball decays when the universe is radiation-dominated, and the
cosmic temperature at the decay is estimated as
\begin{eqnarray}
& & \hspace{-5mm}
T_{\rm D} \simeq \left(\frac{90}{4\pi^2 N_*}\right)^{1/4} 
\sqrt{\Gamma_Q^{\rm (sat)}M_{\rm P}},
\nonumber \\ & & \hspace{0.5mm}
\simeq 2.4~{\rm MeV} 
\left(\frac{M_F}{10^6~{\rm GeV}}\right)^{1/2}
\left(\frac{Q}{10^{23}}\right)^{-5/8},
\label{TD}
\end{eqnarray}
where $N_*$ is the degrees of freedom at the corresponding temperature, and set to be
10.75 here.

On the other hand, the $Q$-ball decay into gravitinos is not saturated. The decay rate
of the elementary process squark $\rightarrow$ quark $+$ gravitino is given by
\begin{equation}
\Gamma_{\phi\rightarrow q \psi_{3/2}} = \frac{1}{48\pi} \frac{m_\phi^5}{m_{3/2}^2M_{\rm P}^2}.
\end{equation}
Since effective coupling is thus estimated as
\begin{equation}
f_{\rm eff} \simeq \frac{1}{\sqrt{6}} \frac{\omega_Q^2}{m_{3/2}M_{\rm P}},
\end{equation}
we have
\begin{equation}
\frac{f_{\rm eff}\phi_Q}{\omega_Q} \simeq \frac{\pi}{\sqrt{3}}\frac{M_F^2}{m_{3/2}M_{\rm P}}
\lesssim \frac{\pi}{\sqrt{3}}\frac{g_s}{(4\pi)^2} \sim 0.01 g_s \ll 1,
\end{equation}
where we use Eqs.(\ref{omegaQ}) and (\ref{phiQ}) in the first equality, and
Eq.(\ref{MFrange}) in the second equality. Therefore, the branching ratio of the decay
into the gravitino is estimated as
\begin{equation}
B_{3/2} \equiv \frac{\Gamma_Q^{(3/2)}}{\Gamma_Q^{\rm (sat)}} \simeq 
3 \pi \frac{f_{\rm eff}\phi_Q}{\omega_Q}
\simeq \sqrt{3} \pi^2 \frac{M_F^2}{m_{3/2}M_{\rm P}}.
\label{B3/2}
\end{equation}

\section{Baryon number, gravitino dark matter, and NLSP abundances}
\label{abundances}
The number densities of the baryon number, the gravitino dark matter, and the NLSP are 
respectively related to the number density of the AD field as
\begin{eqnarray}
& & n_b \simeq \varepsilon b n_\phi, \\
\label{n3/2}
& & n_{3/2} \simeq B_{3/2} n_\phi,\\
\label{nNLSP}
& & n_{\rm NLSP} \simeq \frac{Q_{\rm cr}}{Q} n_\phi.
\end{eqnarray}
WMAP seven-year data tells us $\rho_{\rm DM}/\rho_b=4.94^{+0.95}_{-0.83}$\cite{WMAP}.
Since the ratio of the gravitiono dark matter and the baryon densities is 
\begin{equation}
\frac{\rho_{3/2}}{\rho_b}=\frac{m_{3/2}}{m_N}\frac{n_{3/2}}{n_b} \simeq
\frac{m_{3/2}}{m_N}\frac{B_{3/2}}{\varepsilon b} \simeq 5,
\label{BDM}
\end{equation}
we must have
\begin{equation}
\varepsilon \simeq \frac{\sqrt{3}}{5}\pi^2\frac{M_F^2}{m_{3/2}M_{\rm P}}\frac{m_{3/2}}{b\,m_N}
\simeq 1.4\times 10^{-6} b^{-1} \left(\frac{M_F}{10^6~{\rm GeV}}\right)^2.
\label{epsilon}
\end{equation}
Therefore, the orbit of the AD field should be oblate, which is natural in the gauge-mediated
SUSY breaking models. As mentioned above, this leads to the production of both $Q$ and 
anti-$Q$ balls to cancel out the baryon number to be small enough, compared to
the gravitino dark matter.

Baryon number is created when the AD field starts the rotation. It can be estimated as
\begin{eqnarray}
& & \hspace{-5mm}
Y_b \equiv \frac{n_b}{s} 
\simeq \frac{3}{4} T_{\rm RH} \left.\frac{n_b}{\rho_{\rm rad}} \right|_{\rm RH}
\simeq \frac{3}{4} T_{\rm RH} \left.\frac{n_b}{\rho_{\rm inf}} \right|_{\rm osc}
\nonumber \\ & & \hspace{7mm}
\simeq \frac{3}{4} T_{\rm RH} 
\frac{\varepsilon b m_{\rm eff} \phi_{\rm osc}^2}{3H_{\rm osc}^2M_{\rm P}^2}
\simeq \frac{9}{8\sqrt{2}} \frac{\varepsilon b \phi_{\rm osc}^3 T_{\rm RH}}{M_F^2 M_{\rm P}^2}, 
\nonumber \\ & & \hspace{7mm}
\simeq \frac{9}{8\sqrt{2}} \varepsilon b \beta^{-3/4} \frac{M_F T_{\rm RH}}{M_{\rm P}^2} Q^{3/4}, 
\end{eqnarray}
where $m_{\rm eff} \equiv \sqrt{|V''|} \simeq 2\sqrt{2}M_F^2/\phi_{\rm osc}$ and
$3H_{\rm osc} \simeq m_{\rm eff}$ are used in the last equality in the second line,
and $\phi_{\rm osc} = \beta^{-1/4} M_F Q^{-1/4}$ from Eq.(\ref{Qosc}) is used in the last line.
Inserting Eq.(\ref{epsilon}), we obtain
\begin{eqnarray}
& & \hspace{-5mm}
\left(\frac{Y_b}{10^{-10}}\right) \simeq 0.51 \left(\frac{\beta}{6\times10^{-5}}\right)^{-3/4}
\left(\frac{T_{\rm RH}}{10^6~{\rm GeV}}\right)
\nonumber \\ & & \hspace{21mm} \times 
\left(\frac{Q}{10^{23}}\right)^{3/4}
\left(\frac{M_F}{10^6~{\rm GeV}}\right)^3.
\label{Yb}
\end{eqnarray}
This provides the relation among $Q$, $M_F$, and $T_{\rm RH}$.

On the other hand, the NLSP abundance is estimated as
\begin{eqnarray}
& & \hspace{-5mm}
\frac{\rho_{\rm NLSP}}{s}
= m_{3/2} Y_{3/2} \frac{\rho_{\rm NLSP}}{\rho_{3/2}}, 
\nonumber \\ & & \hspace{6mm}
\simeq 5 m_N Y_b \frac{m_{\rm NLSP} n_{\rm NLSP}}{m_{3/2} n_{3/2}}, 
\nonumber \\ & & \hspace{6mm}
\simeq 5 \frac{1024\pi^2}{81\sqrt{3}} m_N Y_b Q^{-1} \frac{M_F^2M_P}{m_{\rm NLSP}^3},
\nonumber \\ & & \hspace{6mm}
\simeq 3.2\times 10^{-8}~{\rm GeV} \left(\frac{Y_b}{10^{-10}}\right)
\left(\frac{Q}{10^{23}}\right)^{-1}
\nonumber \\ & & \hspace{16mm} \times
\left(\frac{M_F}{10^6~{\rm GeV}}\right)^2
\left(\frac{m_{\rm NLSP}}{300~{\rm GeV}}\right)^{-3},
\label{YNLSP}
\end{eqnarray}
where $Y_{3/2}\equiv n_{3/2}/s$,  and we use $\rho_{3/2}/\rho_b\simeq 5$ in the 
second equality, and Eqs.(\ref{n3/2}), (\ref{nNLSP}), (\ref{Qcr}), and (\ref{B3/2}) in the 
third one. We can now consider the limits of the abundance from BBN constraints
in the next section.

\section{BBN constraints}
\label{BBNconstraints}
\begin{figure}[h]
\begin{tabular}{c}
\includegraphics[width=90mm]{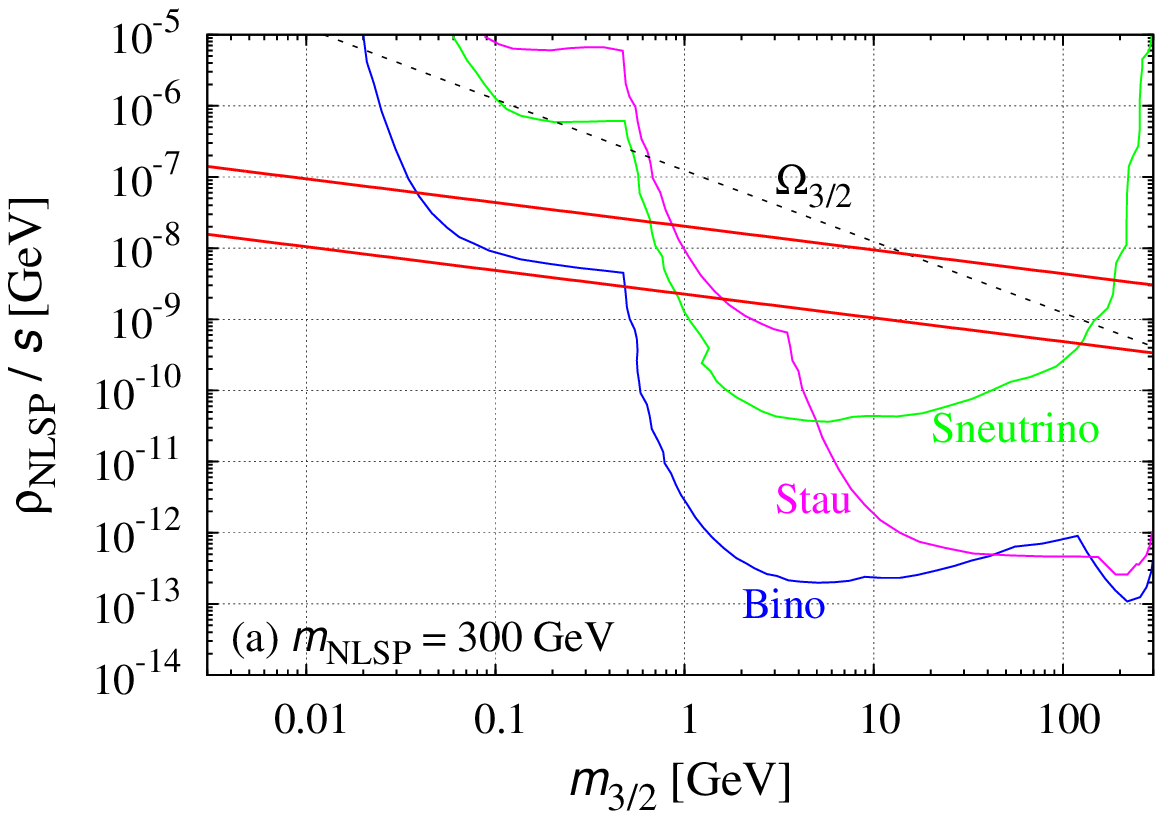} \\
\includegraphics[width=90mm]{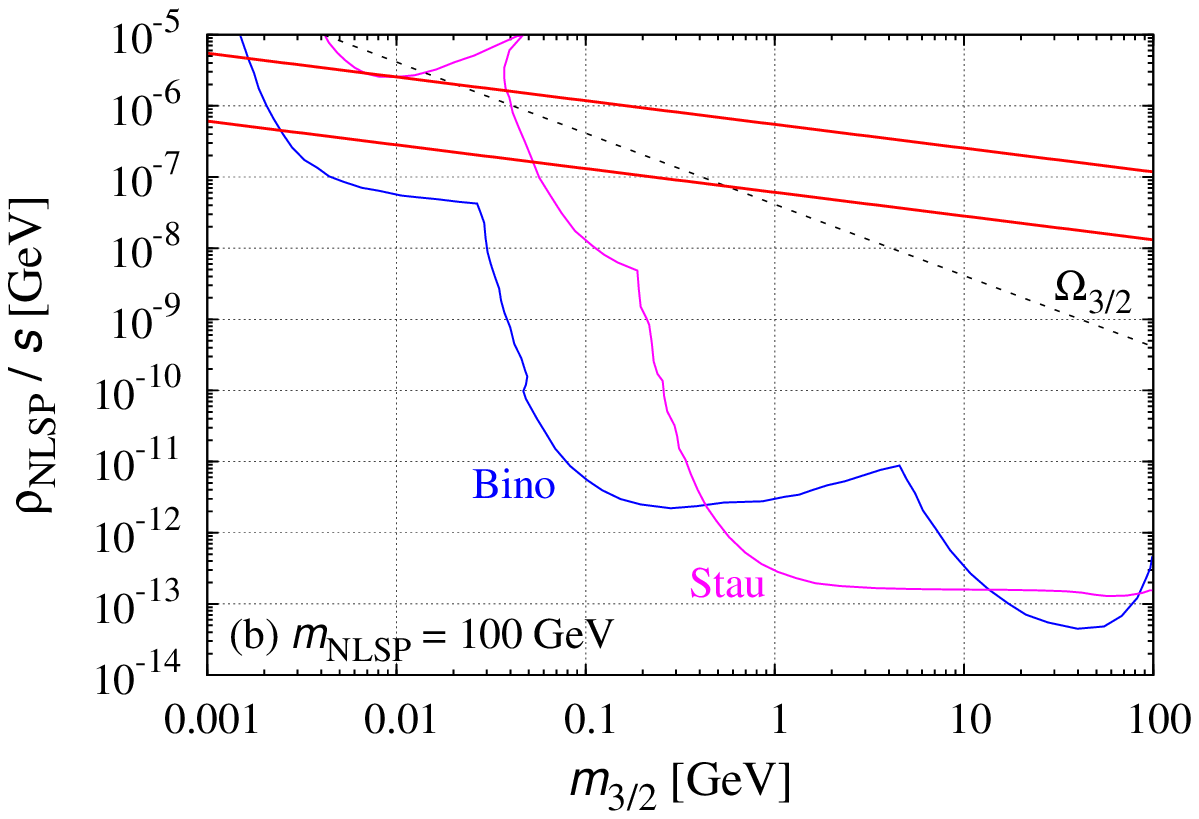} 
\end{tabular}
\caption{NLSP abundance Eq.(\ref{YNLSP1}) for $T_{\rm D}=1$ (3) MeV, denoted as the 
lower (upper) red line, for (a) $m_{\rm NLSP}=300$ GeV and (b) $m_{\rm NLSP}=100$ GeV. 
BBN bounds are shown for bino (blue), stau (pink), and sneutrino (green, only in (a)), 
taken from Ref.\cite{BBN}. Dotted line represents the upper bound of the abundance 
of NSLP that decays into gravitinos, above which overclose the universe [Eq.(\ref{NLSP3/2})]. 
The allowed abundance is in between the red line and BBN bounds or the dotted line.}
\label{fig_NLSP}
\end{figure}

The abundance of the NLSP is limited by BBN constraints, since the decay of the
NLSP affects the abundances of light elements. It depends on the gravitino mass and
the species of the NLSP how it has an influence on them. Let us first estimate the 
expected abundance of the NLSP in the limiting case, which means that the $Q$ ball
decays just in time for BBN ($T_{\rm D}=$1 or 3 MeV), and the thermally produced gravitinos 
do not overdominate. The latter condition is rephrased in terms of the constraint on the
reheating temperature as\cite{KTY}
\begin{equation}
T_{\rm RH} \lesssim 3 \times 10^7~{\rm GeV} 
\left(\frac{m_{\tilde{g}_3}}{500~{\rm GeV}}\right)^{-2}
\left(\frac{m_{3/2}}{\rm GeV}\right),
\label{TRH}
\end{equation}
where $m_{\tilde{g}_3}$ is a gluino mass evaluated at $T=T_{\rm RH}$. 
Using Eqs.(\ref{TD}) and (\ref{Yb}), we can eliminate $M_F$ and $Q$ from 
Eq.(\ref{YNLSP}), which results in
\begin{eqnarray}
& & \hspace{-5mm}
\frac{\rho_{\rm NLSP}}{s}\simeq 7.0 \times 10^{-9}~{\rm GeV} 
\left(\frac{T_{\rm RH}}{10^6~{\rm GeV}}\right)^{-1/3}
\nonumber \\ & & \hspace{20mm} \times
\left(\frac{T_{\rm D}}{\rm MeV}\right)^2
\left(\frac{m_{\rm NLSP}}{300~{\rm GeV}}\right)^{-3},
\nonumber \\ & & \hspace{6.5mm}
\gtrsim 2.3 \times 10^{-9}~{\rm GeV} 
\left(\frac{m_{3/2}}{\rm GeV}\right)^{-1/3}
\nonumber \\ & & \hspace{20mm} \times
\left(\frac{T_{\rm D}}{\rm MeV}\right)^2
\left(\frac{m_{\rm NLSP}}{300~{\rm GeV}}\right)^{-3},
\label{YNLSP1}
\end{eqnarray}
where Eq.(\ref{TRH}) is used in the last inequality.
In Fig.\ref{fig_NLSP}, it is shown in red lines, above which
is allowed. Also shown are the BBN constraints for bino (blue), stau (pink), and 
sneutrino (green) NLSPs, where upper right regions are excluded~\cite{BBN}. In addition, 
we plot the dotted line as the NLSP abundance that the gravitinos produced by the NLSP
decays has a right amount of dark matter:
\begin{eqnarray}
& & \hspace{-5mm}
\left.\frac{\rho_{\rm NLSP}}{s}\right|_{\rm max}
= m_{\rm NLSP} Y_{3/2} 
\nonumber \\ & & \hspace{5mm}
= \frac{m_{\rm NLSP}}{m_{3/2}} m_{3/2} Y_{3/2} 
\simeq \frac{m_{\rm NLSP}}{m_{3/2}} 5 m_N Y_b 
\nonumber \\ & & \hspace{5mm}
\simeq 1.5\times 10^{-7}~{\rm GeV} 
\left(\frac{m_{\rm NLSP}}{300~{\rm GeV}}\right) 
\nonumber \\ & & \hspace{25mm} \times
\left(\frac{m_{3/2}}{\rm GeV}\right)^{-1} 
\left(\frac{Y_b}{10^{-10}}\right),
\label{NLSP3/2}
\end{eqnarray}

Therefore, the NLSP abundance should be larger than the limit Eq.(\ref{YNLSP1}), and
smaller than either excluded regions by BBN constraints in each NLSP species, or the upper
limit that the produced gravitinos from the NLSP decay do not overclose the universe 
[Eq.(\ref{NLSP3/2})]. 
One can see that the gravitino mass above GeV is excluded. The allowed region is
typically $m_{3/2} \lesssim$ GeV for $m_{\rm NLSP} = 300$ GeV,
and $m_{3/2} \lesssim 10^{-2}$ GeV for $m_{\rm NLSP} = 100$ GeV. 
The detail will be found in Table~\ref{table}.

\begin{table}[htdp]
\caption{Upper limit of $m_{3/2}$ in GeV.}
\begin{center}
\begin{tabular}{|c|c|c|c|}
\hline
& & $T_D=1$ MeV & $T_D=3$ MeV \\
\hline
300 GeV& bino & 0.48 & $3.9\times 10^{-2}$ \\
\hline
300 GeV& stau & 1.6 & 0.86 \\
\hline
300 GeV& sneutrino & 0.91 & 0.66 \\
\hline
100 GeV& bino & $2.5\times 10^{-3}$ & $1.9\times 10^{-3}$ \\
\hline
100 GeV& stau & $5.3\times 10^{-2}$ & $2.1\times 10^{-2}$ \\
\hline
\end{tabular}
\end{center}
\label{table}
\end{table}

\section{Successful scenario}
\label{scenario}
As shown in the previous section, the allowed parameter space is restricted by BBN
constraints severely, but there is still a good chance to have successful scenario.
Here we estimate the allowed region in $Q$-ball parameters, namely in $Q-M_F$ plane.

Equation (\ref{Yb}) tells us the iso-$T_{\rm RH}$ lines in $Q-M_F$ plane, which reads
\begin{equation}
Q \simeq 2.4\times 10^{23} 
\left(\frac{T_{\rm RH}}{10^6~{\rm GeV}}\right)^{-4/3}
\left(\frac{M_F}{10^6~{\rm GeV}}\right)^{-4}.
\label{QMF_TRH}
\end{equation}
In the same manner, we can obtain iso-NLSP density lines from Eq.(\ref{YNLSP}) as
\begin{eqnarray}
Q & \simeq & 3.2 \times 10^{23} 
\left(\frac{\rho_{\rm NLSP}/s}{10^{-8}~{\rm GeV}}\right)^{-1}
\nonumber \\ & & \hspace{10mm} \times
\left(\frac{M_F}{10^6~{\rm GeV}}\right)^2
\left(\frac{m_{\rm NLSP}}{300~{\rm GeV}}\right)^{-3},
\label{QMF_YNLSP}
\end{eqnarray}
and iso-$T_{\rm D}$ lines from Eq.(\ref{TD}) as
\begin{equation}
Q\simeq 4.0\times 10^{23}
\left(\frac{T_{\rm D}}{\rm MeV}\right)^{-8/5}
\left(\frac{M_F}{10^6~{\rm GeV}}\right)^{4/5}.
\label{QMF_TD}
\end{equation}

Upper limit of the reheating temperature depends on the gravitino mass $m_{3/2}$
in such a way as in Eq.(\ref{TRH}), and once we fixed $m_{3/2}$ we can restrict 
the NLSP abundances from Fig.\ref{fig_NLSP}.
Therefore, for the fixed $m_{3/2}$, the allowed region is above the line (\ref{QMF_TRH})
with the largest possible $T_{\rm RH}$, and between the lines (\ref{QMF_YNLSP}) 
with the lower and upper NLSP abundances. In addition, the $Q$ ball must decay
before the BBN time so that the temperature at the
decay $T_{\rm D}$ should be larger than 1 MeV (or 3 MeV for more conservative limit),
represented as line (\ref{QMF_TD}).

\begin{figure*}[h]
\begin{tabular}{cc}
\includegraphics[width=90mm]{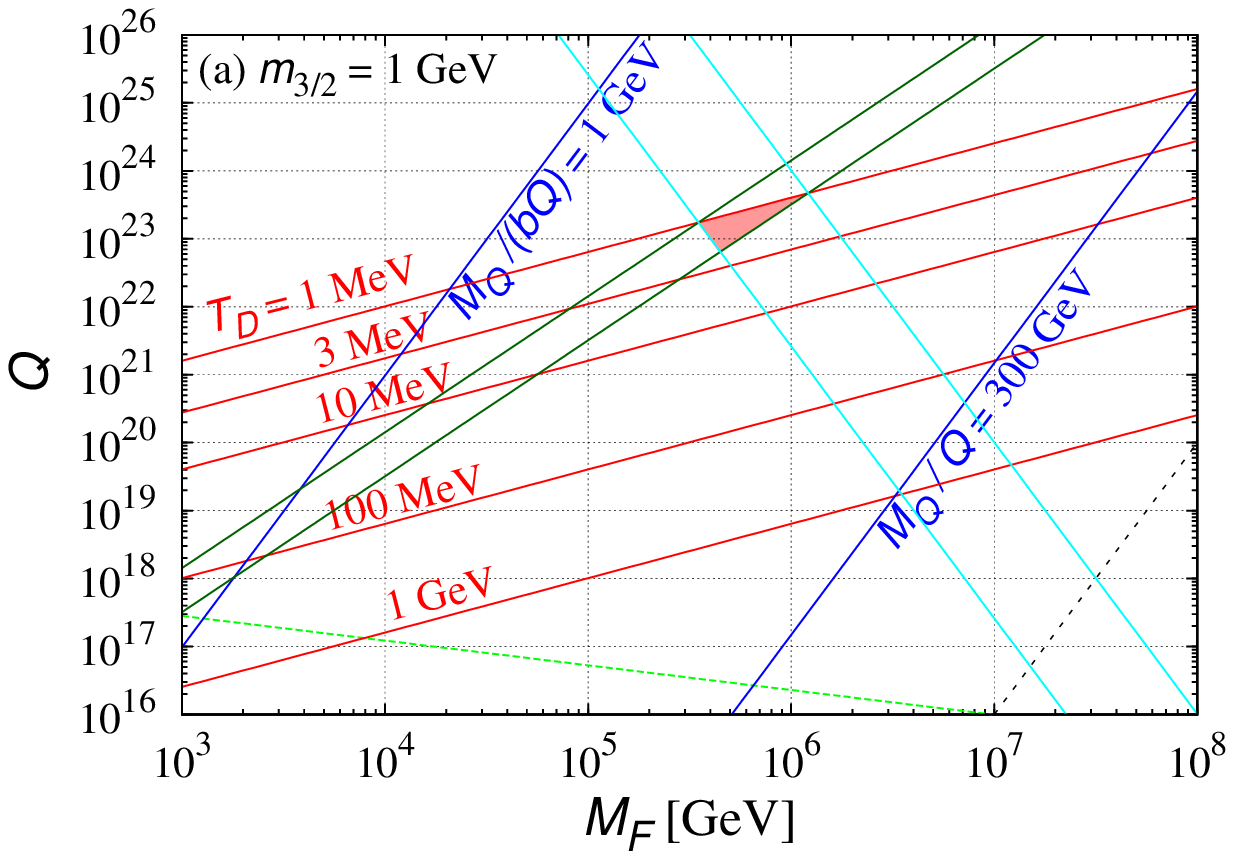} &
\includegraphics[width=90mm]{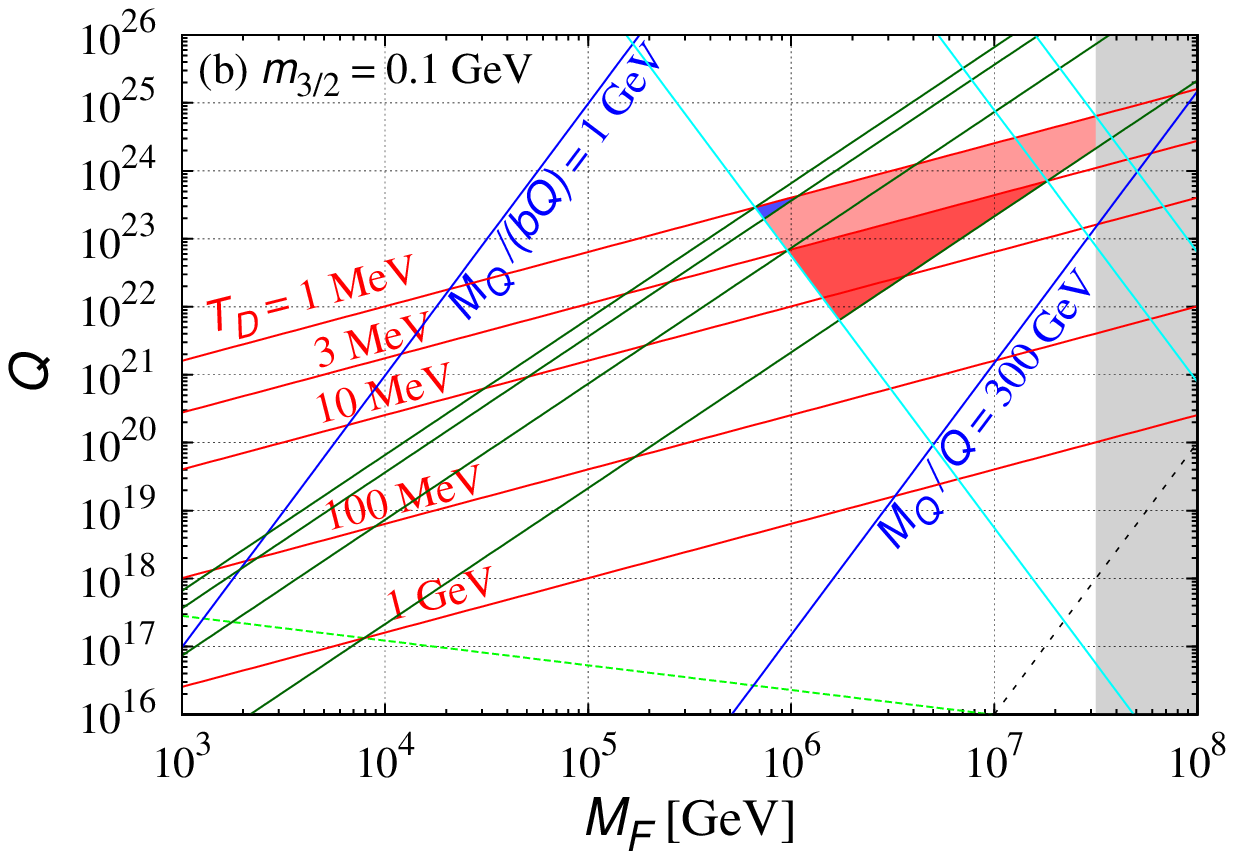} \\
\includegraphics[width=90mm]{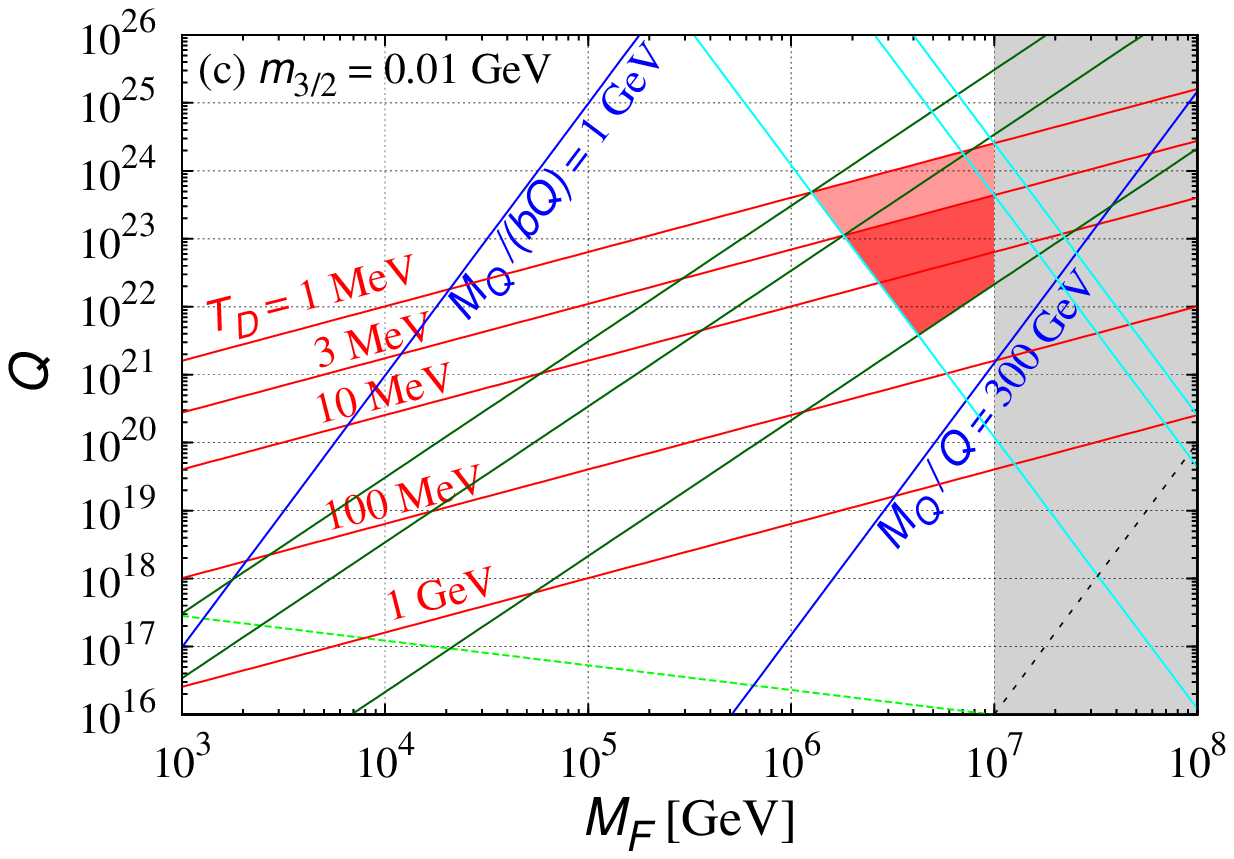} &
\includegraphics[width=90mm]{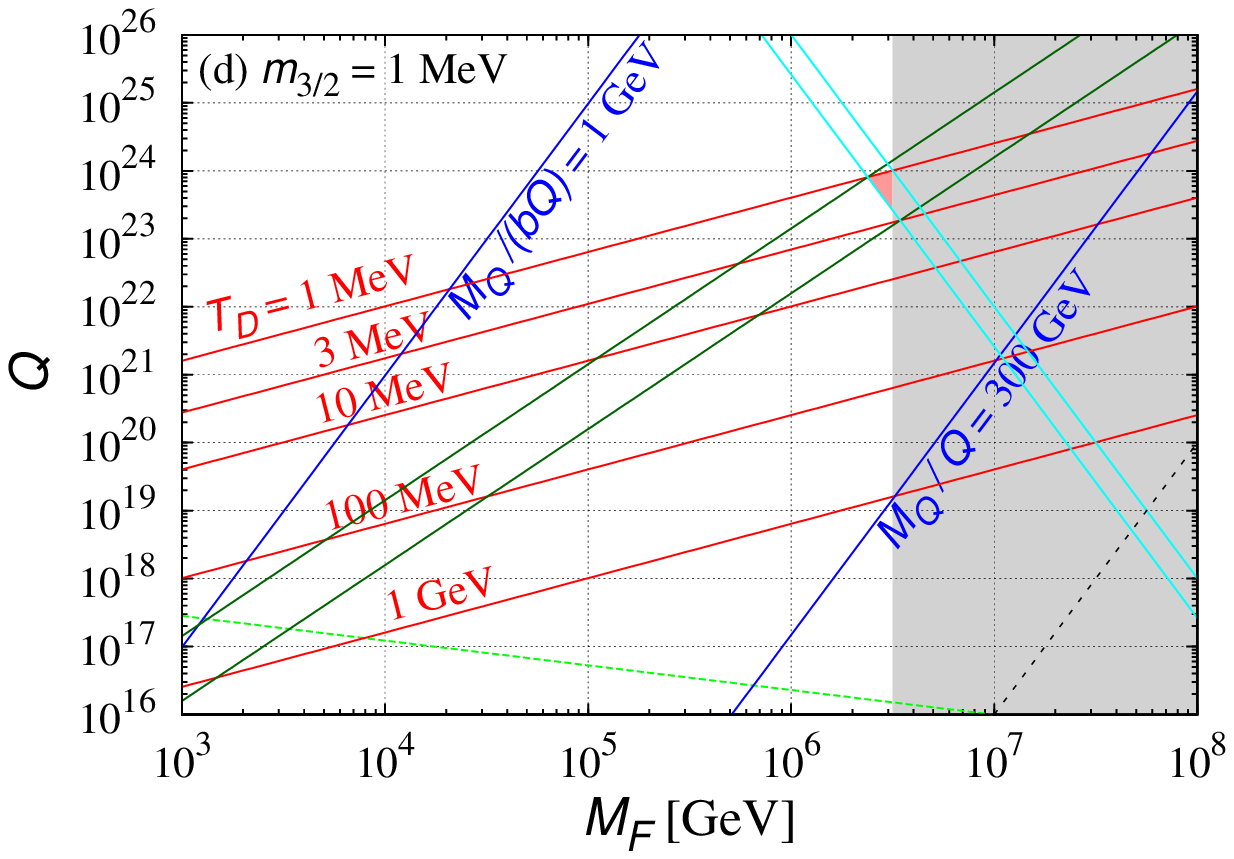} 
\end{tabular}
\caption{Allowed region in $Q-M_F$ plane for (a) $m_{3/2}=1$ GeV, (b) $m_{3/2}=0.1$ GeV,
(c) $m_{3/2}=0.01$ GeV, and (d) $m_{3/2}=1$ MeV. In all figures,
red lines denote iso-$T_{\rm D}$ contours from 1 MeV to 1 GeV from the top to the bottom,
respectively, and blue lines denote the limit of the mass of the particle which the $Q$ ball can
decay into: the upper is for nucleons and the lower is for the 300 GeV NLSP. Here we take
$b=1/3$. Gray hatched region is out of the range of $M_F$ [Eq.(\ref{MFrange})].
\\
(a) Only the stau NLSP has allowed region. Light blue lines
represent the $T_{\rm RH}$ contours for $T_{\rm RH}=3\times 10^7$ GeV (the lower) and 
$3.5\times 10^5$ GeV (the upper). The upper and lower dark green lines indicate 
$\rho_{\rm NLSP}/s=2.3 \times 10^{-9}$ GeV and $1.0\times 10^{-8}$ GeV, 
respectively. 
\\
(b) The faint and the dark red regions are 
applicable to the stau and the sneutrino NLSPs for $T_{\rm D} >1$ MeV and 
$T_{\rm D} >3$ MeV, respectively, while the bino NLSP is only allowed in the blue region.
Light blue lines represent the $T_{\rm RH}$ contours for $T_{\rm RH}=3\times 10^6$ GeV 
(the lower), 74 GeV (the middle), and 2.7 GeV (the upper). The dark green lines indicate 
$\rho_{\rm NLSP}/s=4.9 \times 10^{-9}$ GeV, $8.8\times 10^{-9}$ GeV, 
$4.4\times 10^{-8}$ GeV, and $1.5\times 10^{-6}$ GeV, from the top to the bottom, 
respectively.
\\
(c) The faint and the dark red regions are applicable to all the NLSP species 
for $T_{\rm D} >1$ MeV and $T_{\rm D} >3$ MeV, respectively.
Light blue lines represent the $T_{\rm RH}$ contours for $T_{\rm RH}=3\times 10^5$ GeV 
(the lower), 64 GeV (the middle), and 17 GeV (the upper). The dark green lines indicate 
$\rho_{\rm NLSP}/s=1.0 \times 10^{-8}$ GeV, $9.4\times 10^{-8}$ GeV, and 
$1.5\times 10^{-5}$ GeV, from the top to the bottom, respectively.
\\
(d) Allowed region is appicable to all the NLSP species. Light blue lines represent 
the $T_{\rm RH}$ contours for $T_{\rm RH}=3\times 10^4$ GeV 
(the lower) and $1\times 10^4$ GeV (the upper). The upper and lower dark green 
lines indicate  $\rho_{\rm NLSP}/s=2.3 \times 10^{-8}$ GeV and $2.0\times 10^{-7}$ GeV, 
respectively. 
\\
In all figures, green dashed line shows the limit of the charge evaporated in thermal bath, 
and $Q$ ball is the gauge-mediation type above the black dotted line.}
\label{fig_QMF}
\end{figure*}

We plot these lines for $m_{\rm NLSP}=300$ GeV in $Q- M_F$ planes in 
Fig.\ref{fig_QMF} for $m_{3/2}=$ 1 GeV, 0.1 GeV, 0.01 GeV, and 1 MeV. 
Red lines denote the $T_{\rm D}$ contours, light blue lines are the $T_{\rm RH}$ contours, 
and dark green lines represent the NLSP abundance contours. Also shown in blue are 
the mass per charge $M_Q/Q$ lines:
\begin{equation}
Q \simeq 1.2\times 10^{27} 
\left(\frac{M_F}{10^6~{\rm GeV}}\right)^4
\left(\frac{M_Q/Q}{\rm GeV}\right)^{-4},
\end{equation}
above which is prohibited for the $Q$ ball to decay into those particles with mass
larger than $M_Q/Q$. In particular, the $Q$-ball decay with baryon number is stable 
above the upper blue line denoted by $M_Q/(b\,Q)=1$ GeV [equivalent to Eq.(\ref{QD})], 
and the 300 GeV NLSP is only produced once the charge of the $Q$ ball becomes 
below the lower line denoted as $M_Q/Q=300$ GeV [equivalent to Eq.(\ref{Qcr})].

Only the stau NLSP is allowed in the region shown in faint red in Fig.\ref{fig_QMF}(a). 
In Fig.\ref{fig_QMF}(b), there are three territories, where the faint ($T_{\rm D} > 1$ MeV) 
and the dark ($T_{\rm D} > 3$ MeV) red are applicable to the stau and the sneutrino NLSPs, 
while the blue is for the bino NLSP. On the other hand, all three species (bino, stau, and 
sneutrino NLSPs) are relevant for the allowed region in Fig.\ref{fig_QMF}(c), which
corresponds to the NLSP abundance between the red lines [Eq.(\ref{YNLSP1})] and 
the dotted line [Eq.(\ref{NLSP3/2})] in Fig.\ref{fig_NLSP}(a).
We can see in Fig.~\ref{fig_QMF}(d) where $m_{3/2}=1$ MeV that the allowed region 
disappear completely for $T_{\rm D}>3$ MeV, and only a tiny area remains for 
$T_{\rm D}>1$ MeV. Thus, we obtain the lower bound  of the gravitino mass that could 
result in successful scenario as $m_{3/2} \gtrsim 1.1\times 10^{-3} (6.9\times 10^{-4})$ GeV 
for $T_{\rm D} > 3$ (1) MeV. 

Notice that the allowed regions reside in such parameter space that the $Q$-ball charge
is by far large enough so that the $Q$ ball survives from the evaporation in thermal bath,
where the $Q$ ball evaporates if the charge is less than \cite{KK3}
\begin{equation}
Q_{\rm evap} \simeq 2.3\times 10^{16} 
\left(\frac{M_F}{10^6~{\rm GeV}}\right)^{-4/11}
\left(\frac{m_\phi}{\rm TeV}\right)^{-8/11},
\end{equation}
which is shown by green dashed lines in Fig.\ref{fig_QMF}. On the other hand, 
black dotted lines in these figures denote the boundary above which the $Q$ ball 
is the gauge-mediation type:
\begin{equation}
Q_{\rm gauge} \simeq 10^{12} 
\left(\frac{M_F}{10^6~{\rm GeV}}\right)^4
\left(\frac{m_\phi}{\rm TeV}\right)^{-4}.
\end{equation}

Also notice that those $Q$ balls in the allowed region in Fig.~\ref{fig_QMF}(a)  
($m_{3/2}=1$ GeV) will form when the field starts rotation at the field amplitude
larger than $\sim M_F^2/m_{3/2}$, where the gravity-mediation effects dominate over
the gauge-mediation ones. Depending on the one-loop potential, the formed $Q$ ball
could be the delayed type \cite{KK3} or the new type \cite{New} which later
transforms into the gauge-mediation type, or the gauge-mediation type may form directly.

\section{Conclusions}
\label{conclusion}
We have investigated the scenario of simultaneous production of the baryon asymmetry
and the dark matter of the universe through the $Q$-ball decay in the gauge-mediated
SUSY breaking model. This is simply achieved by the $Q$ ball with charge small enough 
to decay into nucleons to create baryon number of the universe. We have calculated
the branching ratio of the gravitino to find that it is small. Therefore, in order to obtain
the observed baryon to dark matter abundance ratio, the AD field should
rotate in the oblate orbit so that both $Q$ and anti-$Q$ balls are created to suppress
the baryon number compared to the gravitino dark matter. 

At the same time, the charge should be large enough to kinematically forbid the decay 
into NLSPs. This evades the serious BBN constraints on the NLSP abundance, which 
was not properly taken into account in the literatures. NLSPs are actually produced at 
the very end of the $Q$-ball decay when the charge becomes small enough to open 
the corresponding decay channel.

Including all these considerations, we have used the BBN constraints for the bino, stau,
and sneutrino NLSPs in Ref.\cite{BBN}, to get the allowed range of the gravitiono mass.
We have found that typically $m_{3/2} \lesssim 1 \ (10^{-2})$ GeV for 
$m_{\rm NLSP}=300 \ (100)$ GeV. 

We have also shown the allowed region in $Q-M_F$ plane for $m_{\rm NLSP}=300$ GeV.
We thus have found the production of right amounts of both the baryon asymmetry and the 
gravitino dark matter of the universe is successful when the $Q$ ball has the charge 
$Q \simeq 10^{23}$ for $M_F \simeq 10^6 - 10^7$ GeV and 
$m_{3/2} \simeq 10^{-2} - 10^{-1}$ GeV. 
It is naturally achieved, for example, for the $n=6$ $udd$ direction. In this case,
the nonrenormalizable superpotential $W_{\rm NR}= (udd)^2/M_{\rm P}^3$ determines 
the field amplitude at  the onset of the oscillation 
$\phi_{\rm osc} \simeq (M_F^2M_{\rm P}^3)^{1/5}$, and
also leads to the $A$-term of the form $V_A \sim a m_{3/2} W_{\rm NR} + {\rm h.c.}$,
and the scenario is successful for $M_F\sim10^7$ GeV, $m_{3/2}\sim 0.01$ GeV, and
$a \sim 0.1$, which leads to $\phi_{\rm osc} \sim 7\times 10^{13}$ GeV, and $Q \sim 10^{23}$.

\section*{Acknowledgments}
The work is supported by Grant-in-Aid for Scientific Research  
23740206 (S.K.), 14102004 (M.K.) and 21111006 (M.K.) 
from the Ministry of Education, Culture, Sports, Science and 
Technology in Japan, 
and also by World Premier
International Research Center Initiative (WPI Initiative), MEXT, Japan.



\end{document}